 \documentclass[final,3p,times]{elsarticle}

\usepackage{hyperref}
\usepackage{subcaption}
\usepackage{amsmath}

\usepackage{caption}
\usepackage{subcaption}
\usepackage{graphicx}

\hypersetup{
    colorlinks=true,
    pdfborder={0 0 0},
}

\usepackage{amssymb}

\journal{European Journal of Control}

\begin{document}

\begin{frontmatter}



\title{An analysis of $\mathbb{P}$-invariance and dynamical compensation properties from a control perspective}




\author[a]{Akram Ashyani, PhD}
\author[a]{Yu-Heng Wu}
\author[a]{Huan-Wei Hsu}
\author[a,*]{Torbj{\"o}rn E. M. Nordling, PhD} 

\affiliation[a]{{Department of Mechanical Engineering}, {National Cheng Kung University}, {No. 1 University Rd.}, {Tainan} {701}, {Taiwan}}
\affiliation[*]{Corresponding author: torbjorn.nordling@nordlinglab.org}

\begin{abstract}
Dynamical compensation (DC) provides robustness to parameter fluctuations. As an example, DC enable control of the functional mass of endocrine or neuronal tissue essential for controlling blood glucose by insulin through a nonlinear feedback loop. 
Researchers have shown that DC is related to structural unidentifiability and $\mathbb{P}$-invariance property, and $\mathbb{P}$-invariance property is a sufficient and necessary condition for the DC property. 
In this article, we discuss DC and $\mathbb{P}$-invariancy from an adaptive control perspective. 
An adaptive controller is a self-tuning controller used to compensate for changes in a dynamical system. 
To design an adaptive controller with the DC property, it is easier to start with a two-dimensional dynamical model. 
We introduce a simplified system of ordinary differential equations (ODEs) with the DC property and extend it to a general form. 
The value of the ideal adaptive control lies in developing methods to synthesize DC to variations in multiple parameters.
Then we investigate the stability of the system with time-varying input and disturbance signals, with a focus on the system's $\mathbb{P}$-invariance properties. 
This study provides phase portraits and step-like response graphs to visualize the system's behavior and stability properties.
\end{abstract}




\end{frontmatter}



\section{Introduction}
Dynamical compensation (DC) implies that the output of a system does not depend on a parameter for any input \citep{Karin2016}. 
For instance, in glucose homeostasis controlled by insulin, despite parameter variations, the glucose response remains identical. 
This definition of the DC property is a sufficient condition and implies that the parameter is structurally unidentifiable \citep{Villaverde2017Dynamic,Villaverde2017DynamicalAndDtructuralIdentifiability,Villaverde2022}. 
In 2017, a necessary and sufficient condition for the DC property was introduced using equivariances and partial differential equations, denoted as the $\mathbb{P}$-invariance property \citep{Sontag2017}. 
The $\mathbb{P}$-invariance property related to a parameter indicates that changing the parameter does not alter the system's behavior, which is useful in biological and medical models. 
This phenomenon is especially advantageous when a change in a parameter has no effect on the output, allowing the system's behavior to be predicted. 

Robustness, which refers to a system's ability to handle fluctuations, is critical in dynamical systems. 
Several studies on adaptation and homeostasis have demonstrated the robustness of biological systems, such as the robustness of bacterial chemotaxis \citep{Barkai:1997:Robustness-in-simple-biochemical-networks.:oa,Alon1999}. 
The application of DC and $\mathbb{P}$-invariance properties is also beneficial in epidemiological models \citep{Sauer2021,Browning2020}. 
Therefore, the DC property may be included in future robustness research. 
Karin et al. used the glucose homeostasis model to discuss the robustness and DC property of homeostasis \citep{Karin2016}.

Several mathematical models based on systems of differential equations have been developed to comprehensively analyze biological observations and identify all possible connections \citep{Ashyani2016,Ashyani2016b,Ashyani2018}. 
However, it is often more convenient to work with simpler models with fewer dimensions, as they are easier to interpret and analyze.
In this paper, we aim to simplify the original model in Karin et al. and include another feedback mechanism to derive an extended model in section \ref{mathematical_model}. 
We began by checking the system's stability in section \ref{sec:results} because the system must be stable to check the DC and $\mathbb{P}$-invariance properties. 
We use the phase portrait approach to verify the system's stability and obtain some results, for preferred stable situations, to compare the results of the DC and $\mathbb{P}$-invariance properties. 
Finally, in the numerical simulation in section \ref{Numerical_simulation}, we considered situations in which the system is stable at desired equilibrium points and demonstrate the impact of adaptive control and $\mathbb{P}$-invariance in the system when it is perturbed.

\section{Mathematical model}
\label{mathematical_model}
	In our study, as a starting point, we used the hormonal circuit reactions model stated in Karin et al. \cite{Karin2016};
	\begin{subequations}
\label{eq:dc_original_model}
	\begin{align}
\label{eq:dc_original_model_1}	\frac{dy}{dt} &= u_{0} + u(t) - sx(t)y(t), \\
\label{eq:dc_original_model_2}	\frac{dx}{dt} &= pz(t)y(t)-x(t), \\
\label{eq:dc_original_model_3}	\frac{dz}{dt} &= z(t)(y(t)-y_{0}),
	\end{align}
	\end{subequations}
	where $s$ and $p$ are the feedback gains of $x(t)$ and $z(t)$, respectively.
	The output variable, $y(t)$, is a regulated variable that is able to form a feedback loop with $x(t)$ and $z(t)$.
	The regulated variable $y(t)$ controls the functional mass $z(t)$ of tissue which secretes hormone $x(t)$ in this circuit.
	The aim is to first simplify the model \ref{eq:dc_original_model}, containing Eqs. \ref{eq:dc_original_model_1}-\ref{eq:dc_original_model_3}, then control the system with adaptive proportional-integral feedback so that it has $\mathbb{P}$-invariance property. 
We also compared the differences between DC and $\mathbb{P}$-invariance property. 

	We simplified the model \ref{eq:dc_original_model} as, 
	\begin{subequations}
\label{eq:dc_simplified}
	\begin{align}
	\label{simp1}
	\frac{dy}{dt} &= u_{0} + u(t) - sz(t)y(t), \\
	\label{simp2}
	\frac{dz}{dt} &= z(t)(y(t)-y_{0}),
	\end{align}
	\end{subequations} 
where  $z(t)$ is the feedback state, and $y(t)$ is the output of the system.
	Our expectation is that the positive constant $s$ has the DC property, meaning that the output $y(t)$ is invariant to the change of the parameter $s$. 
	Hence we introduce $\tilde{z}(t) = sz(t)$ and substitute $z(t)$ in Eq. \ref{simp1} and Eq. \ref{simp2} with $\tilde{z}(t)$ resulting in
	\begin{subequations}
		\label{eq:dc_simplified_variable_substitution}
		\begin{align}
		\frac{dy}{dt} &= u_{0} + u(t) - \tilde{z}(t)y(t), \\
		\frac{d\tilde{z}}{dt} &= \tilde{z}(t)(y(t)-y_{0}).
		\end{align}
	\end{subequations}
	The above Equations show that the output response $y(t)$ remains the same when the value of $s$ changes.

\begin{figure}[h]
	\begin{center}
		\includegraphics[width=0.8\columnwidth]{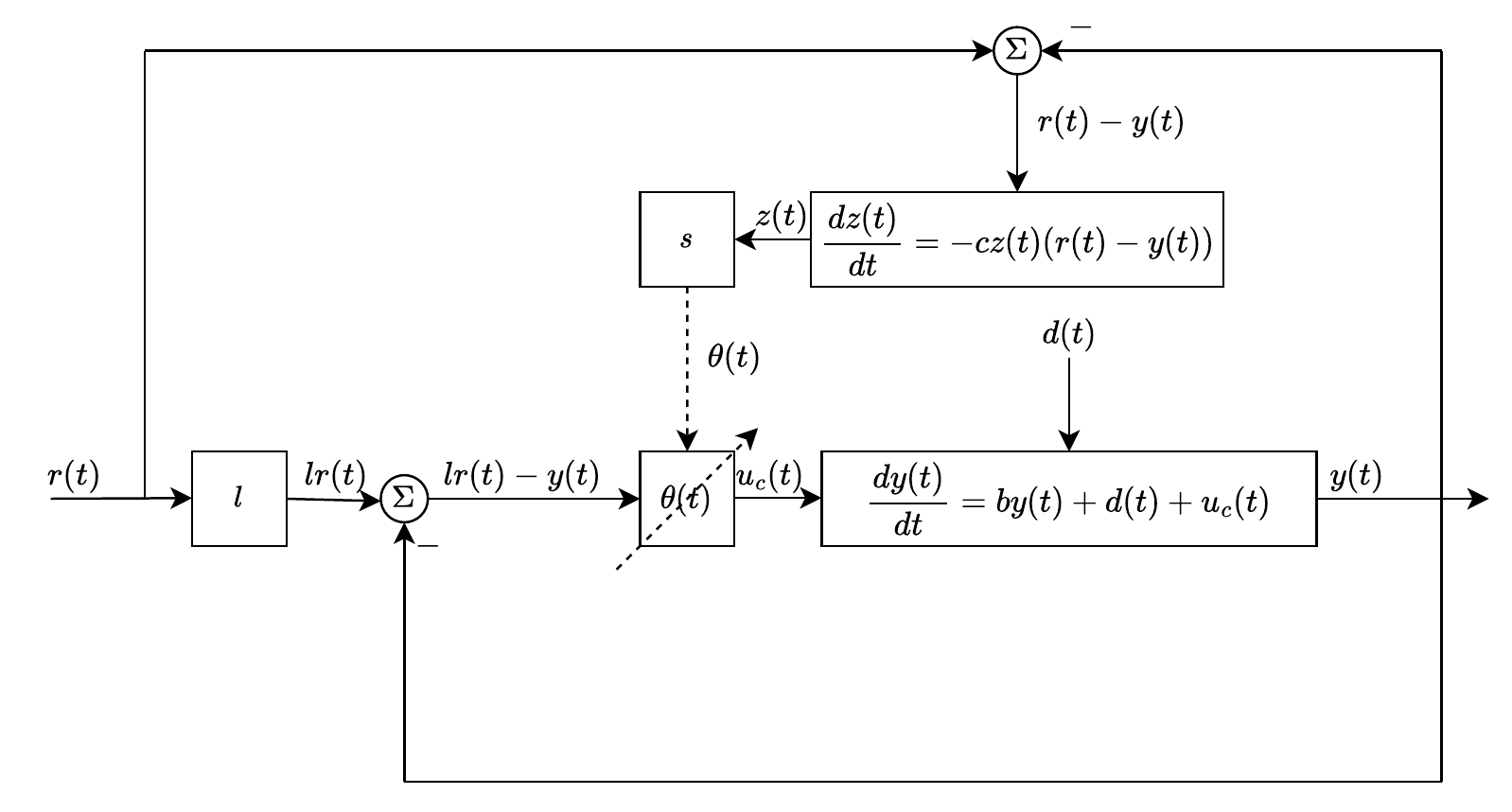}
	\end{center}
	\caption{Block diagram shows the adaptive proportional-integral feedback $ sz(t)\big(lr(t)-y(t)\big)$ where $sz(t)$ is the adaptive proportional-integral gain with two error terms  $r(t)-y(t)$ and $lr(t)-y(t)$. 
Term $d(t)$ represents the disturbance.}
	\label{fig:block_diagram_adaptive_PI}
\end{figure}

By extending our simplified model with a DC property in parameter $s$, we created an adaptive proportional-integral feedback model
	\begin{subequations}\label{extend}
	\begin{align}
		\label{extend1}
		\frac{dy}{dt} &= by(t) + d(t) + sz(t)\big(lr(t)-y(t)\big) , \\
		\label{extend2}
		\frac{dz}{dt} &= -cz(t)\big(r(t)-y(t)\big).
	\end{align}
	\end{subequations}
The system can be viewed as an open-loop exponential growth system $\frac{dy}{dt} = by(t)$, where $d(t)$ and $r(t)$ represent the disturbance and reference input, respectively. 
The error term is given by $r(t)-y(t)$ and $lr(t)-y(t)$, and the adaptive proportional-integral feedback is $sz(t)\big(lr(t)-y(t)\big)$, where $sz(t)$ is considered as the adaptive proportional-integral gain. 
In control theory, a reference input refers to an input signal that guides the system response. 
Typically, the goal is to make the response $y(t)$ track the reference input $r(t)$, such that the error term is zero ($r(t)-y(t)=0$) at the equilibrium point. 
Furthermore, since we consider this equation in the context of biological phenomena, all parameters are assumed positive. 
This implies that $b$, $s$, $l$, and $c$ are all positive, and for every $t>0$, all $y(t)$, $z(t)$, and $r(t)$ are positive. 
The block diagram of the adaptive proportional-integral system is illustrated in Fig. \ref{fig:block_diagram_adaptive_PI}.

To verify the DC property of our model, the system should be at an equilibrium point before being perturbed by any input. 
When a system is at an equilibrium point, its value does not change with time. 
We then triggered the system with a step-like input $r(t)$ to sketch the response $y(t)$. 
We adjust the value of each parameter in Eqs. \ref{extend1} and \ref{extend2} to observe how they affect the system. 
The stability region is discovered by drawing the phase portrait.

	\section{Results}
	\label{sec:results}
Here, we began by checking stability of the system, and then we compared the differences between $\mathbb{P}$-invariance and DC property. 
Finally, we provided a numerical example to illustrate the result.
\subsection{Phase portrait and stability}
\label{append:dc_modified_phase}
Our goal is to discover the region of attraction by drawing the phase portrait. 
Our goal is to discover the region of attraction by drawing the phase portrait. 
By setting the derivative terms in Eqs. \ref{extend1} and \ref{extend2} to zero and assuming that the reference $r(t) = r$ and disturbance $d(t) = d$ remain constant, two equilibrium points can be obtained: 
\begin{align}
\label{equilibrium point1}
&E_1=(y_1,z_1)=\big(-\frac{d}{b},0\big), \\
\label{equilibrium point2}
&E_2=(y_2,z_2)=\big(r,\frac{d+br}{sr(1-l)}\big).
\end{align}
Under the assumption that all parameters are non-negative and the signals $d$ and $r$ are positive, we note the following:
Since $y_1$ is negative in the equilibrium point $E_1$, it is a biologically infeasible state of the system. 
If $0<l \leq 1$, then both $z_2$ and $y_2$ are non-negative, making $E_2$ the equilibrium point of interest.
To ensure that $z_2$ remain finite, we first assume $0<l<1$. 
The local stability of a system can be analyzed by calculating the eigenvalues of the matrix of partial derivatives in equilibrium points, known as the Jacobian matrix. 
The matrix of partial derivatives for system \ref{extend} and its eigenvalues are shown below.
\begin{align}
	\label{eq:dc_modified_Jacobian}
		&\boldsymbol{J}(y,z) = \begin{bmatrix}
	b-sz(t) & s\big(lr-y(t)\big)\\
	cz(t) & -c\big(r-y(t)\big)
	\end{bmatrix}, \\
	&\lambda(y,z) =\frac{1}{2}\Big(
	b-sz(t)-c\big(r-y(t)\big) 
 \pm
\\
\nonumber
& \scriptstyle \sqrt{\Big(b-sz(t)-c\big(r-y(t)\big)\Big)^2-4c\Big(z(t)\big(sr(1-l)\big)-b\big(r-y(t)\big)\Big)} \Big).
\end{align}
The Jacobian matrix is presented in an algebraic structure to calculate the eigenvalues easier when analyzing the local stability of the individual equilibrium point.

{\bf{1) Local stability of $E_1$:}}
To investigate the local stability around $E_1$, we computed two eigenvalues.
\begin{align}
&\lambda_1(y_1,z_1)=b,\hspace{2mm}
\lambda_2(y_1,z_1)= -c\big(d+b r\big)/b.
\end{align}
As $b>0$ and $-c\big(d+br\big)/b<0$ this equilibrium point is a saddle point.

{\bf{2) Local stability of $E_2$:}}
For the equilibrium point $E_2$ the eigenvalues are
\begin{align}
\lambda_{1}=\frac{\tau+\sqrt{\tau^2-4\delta}}{2}, \hspace{5mm}
	\lambda_{2}=\frac{\tau-\sqrt{\tau^2-4\delta}}{2},
\end{align}
where
\begin{align}
\tau&=\mathrm{trace} \big(J(y_2,z_2)\big)=\frac{d+blr}{r(l-1)},\\
\delta&=\det \big(J(y_2,z_2)\big)= c\big(d+br\big).
\end{align}
Three situations can happen:
\begin{enumerate}
\item[1)]$\tau^2-4\delta=0$,
\item[2)]$\tau^2-4\delta<0$,
\item[3)]$\tau^2-4\delta>0$.
\end{enumerate}
In both (1) and (2), stability depends on $\tau$. 
Hence, if $\tau <0$, then $E_2$ is stable. 
Based on the assumption that parameters and variables are positive to be meaningful in biology and $l<1$, we have $\tau <0$, which means that $E_2$ is a stable equilibrium point. 
In situation (3), as $\delta >0$, it will result in $|\tau|>\sqrt{\tau^2-4\delta}$. 
Hence, if $\tau <0$, then $E_2$ is stable. 
Again, based on the assumption of having meaningful parameters in the equilibrium points, $l<1$, the equilibrium $E_2$ is stable. 
These findings demonstrate that as long as $E_2$ is meaningful in biology, it is a globally stable equilibrium point.
\subsection{Influence of the adaptive controller on stability}
\label{subsec:dc_p-invariance_result}
Our aim is to investigate the influence of adaptive controller term on the stability of the system.
	Given a system
	\begin{align}
	\label{eq:dc_pinvariance_def_system}
	\dot{x} = f(x(t),u(t),p),\; y = g(x(t),u(t),p), \; x(0) = \gamma_{p};
	\end{align}
	if there exists an equivalent transformation
	\begin{subequations}
	\label{eq:dc_pinvariance_def}
	\begin{align}
	\label{p_invariancy_f} f(\eta_{p}(x),u,p) &= (\eta_{p})_*(x)f(x,u), \\
	\label{p_invariancy_g} g(\eta_{p}(x),u,p) &= g(x,u), \\
	\label{p_invariancy_eta} \eta_{p}(\gamma) &= \gamma_{p},
	\end{align}
	\end{subequations}
	where $\eta_*$ denotes the Jacobian matrix of transformation $\eta$, the system have the $\mathbb{P}$-invariance property \cite{Sontag2017}.
By verifying the invariance of the system \ref{extend}, containing Eqs. \ref{extend1}-\ref{extend2}, we were able to discover the parameters that lead to the DC property. 
We also demonstrated the differences between the definitions of $\mathbb{P}$-invariance and the DC property. 
Based on the definition of the $\mathbb{P}$-invariance property and the relationship with the DC property, the DC property can be classified as an adaptive control strategy in the system \ref{extend}.

\subsubsection{Verification of the $\mathbb{P}$-invariance property}
	We verify that the system is $\mathbb{P}$-invariant with respect to variation of $s$.
	In order to verify that the system \ref{extend} has the $\mathbb{P}$-invariance property, we introduced $x_{1}(t)$ and $x_{2}(t)$ as two state variables and $y(t)$ as the output variable of the system. 
For simplicity, we wrote the system \ref{extend} in $x_1$, $x_2$, and $y$ as,
	\begin{subequations}\label{eq:dc_extended_control}
		\begin{align}
		\label{eq:dc_extended_x1}
		\dot{x_{1}} &=  -cx_{1}\big(r(t)-x_{2}\big), \\
		\label{eq:dc_extended_x2}
		\dot{x_{2}} &=  bx_{2} + d(t)+sx_{1}\big(lr(t)-x_{2}\big), \\
		\label{eq:dc_extended_y}
		y &= x_{2}.
		\end{align}
	\end{subequations}
	The notation here is selected to be identical to the one used by \cite{Sontag2017}. 
We considered the possible equivariance $\eta_p(x_1,x_2)=\big(\alpha_p(x_1,x_2), \beta_p(x_1,x_2)\big)$. 
In this case, the condition $g\big(\eta_p(x),u,p\big)=g(x,u)$ means $\beta_p(x_1,x_2)=x_2$. 
Therefore, we have $\eta_p(x_1,x_2)=\big(\alpha_p(x_1,x_2), x_2\big)$. 
Hence, 
\begin{align}
(\eta_p)_*(x_1,x_2)&=
\begin{bmatrix}
\frac{\partial \alpha_p}{\partial x_1}(x_1,x_2) & \frac{\partial \alpha_p}{\partial x_2}(x_1,x_2)\\
\frac{\partial x_2}{\partial x_1}&\frac{\partial x_2}{\partial x_2}
\end{bmatrix}
=\begin{bmatrix}
\frac{\partial \alpha_p}{\partial x_1}(x_1,x_2) & \frac{\partial \alpha_p}{\partial x_2}(x_1,x_2)\\
0&1
\end{bmatrix}.
\end{align}
As a result from equation \ref{p_invariancy_f}, for parameter $s$ our aim is to prove 
$ f(\eta_{s}(x),u,s) = (\eta_{s})_*(x)f(x,u)$.
It means:
\begin{align}
\begin{bmatrix}
-c\alpha_{s}(x_{1},x_{2})\big(r(t)-x_{2}\big)  \\
bx_{2}+d(t)+s\alpha_{s}(x_{1},x_{2})\big(lr(t)-x_{2}\big)
\end{bmatrix}
=
\begin{bmatrix}
\frac{\partial \alpha_s}{\partial x_1}(x_1,x_2) & \frac{\partial \alpha_s}{\partial x_2}(x_1,x_2)\\
0&1
\end{bmatrix}
\begin{bmatrix}
-cx_{1}\big(r(t)-x_{2}\big)
\\
bx_{2}+d(t)+x_{1}\big(lr(t)-x_{2}\big)
\end{bmatrix}.
\end{align}
Hence:
	\begin{align}
	\label{p-invariance_x1_k1}
	& -c\alpha_{s}(x_{1},x_{2})\big(r(t)-x_{2}\big) = \frac{\partial\alpha_{s}(x_{1},x_{2})}{\partial x_{1}}\Big(-cx_{1}\big(r(t)-x_{2}\big)\Big) 
+ \frac{\partial\alpha_{s}(x_{1},x_{2})}{\partial x_{2}}\Big(bx_{2}+d(t)+x_{1}\big(lr(t)-x_{2}\big)\Big),	\\
	\label{p-invariance_x2_k1}
		&bx_{2}+d(t)+s\alpha_{s}(x_{1},x_{2})\big(lr(t)-x_{2}\big)=
	bx_{2}+d(t)+x_{1}\big(lr(t)-x_{2}\big).
	\end{align}
By comparing the coefficients in  Eq. \ref{p-invariance_x1_k1} we have
\begin{align}
\frac{\partial\alpha_{s}(x_{1},x_{2})}{\partial x_{1}}&=\frac{\alpha_{s}(x_{1},x_{2})}{x_{1}},\\
\frac{\partial\alpha_{s}(x_{1},x_{2})}{\partial x_{2}}&=0.
\end{align}
From  Eq. \ref{p-invariance_x2_k1} we attained
\begin{align}
s\alpha_{s}(x_{1},x_{2})(lr(t)-x_{2})=x_{1}\big(lr(t)-x_{2}\big),
\end{align}
If $lr(t)-x_{2} \neq 0$, it means 
\begin{align}
\alpha_{s}(x_{1},x_{2})=\frac{x_1}{s}.
\end{align}
Therefore, there is a Jacobian matrix $\alpha_{s}(x_{1},x_{2})= x_1/s$ that can achieve the transformation of the system. 
The system could demonstrate $\mathbb{P}$-invariance when $s$ is the $\mathbb{P}$-invariance parameter.
%

Next, we investigated whether the parameter $b$ has the $\mathbb{P}$-invariance property, meaning that changing $b$ will not influence the behavior of $y(t)$.
As a result from equation \ref{p_invariancy_f}, for parameter $b$ our aim is to prove 
$ f(\eta_{b}(x),u,b) = (\eta_{b})_*(x)f(x,u)$.
It means:
\begin{align}
\begin{bmatrix}
-c\alpha_{b}(x_{1},x_{2})\big(r(t)-x_{2}\big)  \\
bx_{2}+d(t)+s\alpha_{b}(x_{1},x_{2})\big(lr(t)-x_{2}\big)
\end{bmatrix}
=
\begin{bmatrix}
\frac{\partial \alpha_b}{\partial x_1}(x_1,x_2) & \frac{\partial \alpha_b}{\partial x_2}(x_1,x_2)\\
0&1
\end{bmatrix}
\begin{bmatrix}
-cx_{1}\big(r(t)-x_{2}\big)
\\
x_{2}+d(t)+sx_{1}\big(lr(t)-x_{2}\big)
\end{bmatrix}.
\end{align}
	Thus, it is essential to solve
	\begin{align}
	\label{p-invariance_b_1}
		&-c\alpha_{b}(x_{1},x_{2})\big(r(t)-x_{2}\big) = \frac{\partial\alpha_{b}(x_{1},x_{2})}{\partial x_{1}}\Big(-cx_{1}\big(r(t)-x_{2}\big) \Big) 
	+\frac{\partial\alpha_{b}(x_{1},x_{2})}{\partial x_{2}}\Big(x_{2}+d(t)+sx_{1}\big(lr(t)-x_{2}\b)\Big),
	\\
	\label{p-invariance_b_2}
		&bx_{2}+d(t)+s\alpha_{b}(x_{1},x_{2})\big(lr(t)-x_{2}\big)=
	x_{2}+d(t)+sx_{1}\big(lr(t)-x_{2}\big).
	\end{align}
By comparing the coefficients in Eq. \ref{p-invariance_b_1}, we have
\begin{align}
\label{p-invariance_b1}\frac{\partial\alpha_{b}(x_{1},x_{2})}{\partial x_{1}}&=\frac{\alpha_{b}(x_{1},x_{2})}{x_{1}},\\
\label{p-invariance_b2}\frac{\partial\alpha_{b}(x_{1},x_{2})}{\partial x_{2}}&=0.
\end{align}
From  Eq. \ref{p-invariance_b_2} we have
\begin{align}
  bx_2+s\alpha_{b}(x_{1},x_{2})\big(lr(t)-x_{2}\big)=
x_2+sx_{1}\big(lr(t)-x_{2}\big),
\end{align}
If $lr(t)-x_{2} \neq 0$, it means 
\begin{align}\label{b}
\alpha_{b}(x_{1},x_{2})=\frac{x_2+sx_{1}\big(lr(t)-x_{2}\big)-bx_2}{s\big(lr(t)-x_{2}\big)},
\end{align}
and yields
\begin{align}
\label{p-invariance_b1_1}\frac{\partial\alpha_{b}(x_{1},x_{2})}{\partial x_{1}}&=1,\\
\frac{\partial\alpha_{b}(x_{1},x_{2})}{\partial x_{2}}&=\frac{(1-b)\big(slr(t)\big)}{\Big(s\big(lr(t)-x_{2}\big)\Big)^2}.
\end{align}
	There is no solution of $\alpha_{b}(x_{1},x_{2})$ that can be obtained from Eq. \ref{b} and satisfies the two conditions in Eqs. \ref{p-invariance_b1} and \ref{p-invariance_b2}.
	This implies that the system is not $\mathbb{P}$-invariant in $b$.

	Next we verify that the system is not $\mathbb{P}$-invariant with respect to variation of $c$.
	In order to verify that the system \ref{extend} has the $\mathbb{P}$-invariance property, we introduced $x_{1}(t)$ and $x_{2}(t)$ as two state variables and $z(t)$ as the output variable of the system. 
For simplicity, we wrote the system \ref{extend} in $x_1$, $x_2$, and $z$ as,
	\begin{subequations}\label{eq:dc_extended_control}
		\begin{align}
		\label{eq:dc_extended_z1}
		\dot{x_{1}} &=  bx_{1} + d(t)+sx_{2}\big(lr(t)-x_{1}\big), \\
\label{eq:dc_extended_z2}
		\dot{x_{2}} &=  -cx_{2}\big(r(t)-x_{1}\big), \\
		\label{eq:dc_extended_z}
		z &= x_{2}.
		\end{align}
	\end{subequations}
	The notation here is selected to be identical to the one used by \cite{Sontag2017}. 

We considered the possible equivariance $\eta_p(x_1,x_2)=\big(\alpha_p(x_1,x_2), \beta_p(x_1,x_2)\big)$. 
In this case, the condition $g\big(\eta_p(x),u,p\big)=g(x,u)$ means $\beta_p(x_1,x_2)=x_2$. 
Therefore, we have $\eta_p(x_1,x_2)=\big(\alpha_p(x_1,x_2), x_2\big)$. 
Hence, 
\begin{align}
(\eta_p)_*(x_1,x_2)&=
\begin{bmatrix}
\frac{\partial \alpha_p}{\partial x_1}(x_1,x_2) & \frac{\partial \alpha_p}{\partial x_2}(x_1,x_2)\\
\frac{\partial x_2}{\partial x_1}&\frac{\partial x_2}{\partial x_2}
\end{bmatrix}
=\begin{bmatrix}
\frac{\partial \alpha_p}{\partial x_1}(x_1,x_2) & \frac{\partial \alpha_p}{\partial x_2}(x_1,x_2)\\
0&1
\end{bmatrix}.
\end{align}
As a result from equation \ref{p_invariancy_f}, for parameter $c$ our aim is to prove 
$ f(\eta_{c}(x),u,c) = (\eta_{c})_*(x)f(x,u)$.
It means:
\begin{align}
& \nonumber
\begin{bmatrix}
b\alpha_{c}(x_{1},x_{2}) + d(t)+sx_{2}\big(lr(t)-\alpha_{c}(x_{1},x_{2})\big)\\
-cx_{2}\big(r(t)-\alpha_{c}(x_{1},x_{2})\big)
\end{bmatrix}
=  \\
&
\begin{bmatrix}
\frac{\partial \alpha_c}{\partial x_1}(x_1,x_2) & \frac{\partial \alpha_c}{\partial x_2}(x_1,x_2)\\
0&1
\end{bmatrix}
\begin{bmatrix}
bx_{1} + d(t)+sx_{2}\big(lr(t)-x_{1}\big)
\\
-x_{2}\big(r(t)-x_{1}\big)
\end{bmatrix}.
\end{align}
Hence:
	\begin{align}
	\nonumber & b\alpha_{c}(x_{1},x_{2}) + d(t)+sx_{2}\big(lr(t)-\alpha_{c}(x_{1},x_{2})\big) = \frac{\partial\alpha_{c}(x_{1},x_{2})}{\partial x_{1}}\Big(bx_{1} + d(t)+sx_{2}\big(lr(t)-x_{1}\big)\Big) \\
\label{p-invariance_x1_c} &
+ \frac{\partial\alpha_{c}(x_{1},x_{2})}{\partial x_{2}}\Big(-x_{2}\big(r(t)-x_{1}\big)\Big),	\\
	\label{p-invariance_x2_c}
		&-cx_{2}\big(r(t)-\alpha_{c}(x_{1},x_{2})\big)=
	-x_{2}\big(r(t)-x_{1}\big).
	\end{align}
By comparing the coefficients in  Eq. \ref{p-invariance_x1_c} we have
\begin{align}
\label{p-invariance_c1}\frac{\partial\alpha_{c}(x_{1},x_{2})}{\partial x_{1}}&=\frac{b\alpha_{c}(x_{1},x_{2}) + d(t)+sx_{2}\big(lr(t)-\alpha_{c}(x_{1},x_{2})\big)}{bx_{1} + d(t)+sx_{2}\big(lr(t)-x_{1}\big)},\\
\label{p-invariance_c2}\frac{\partial\alpha_{c}(x_{1},x_{2})}{\partial x_{2}}&=0.
\end{align}
From  Eq. \ref{p-invariance_x2_c} we attained
\begin{align}
c x_2 \alpha_{c}(x_{1},x_{2})=x_{2}\big((c-1)r(t)+x_1),
\end{align}
If $cx_{2} \neq 0$, it means 
\begin{align}
\label{c1} \alpha_{c}(x_{1},x_{2})=\frac{(c-1)r(t)+x_1}{c},
\end{align}
and yields
\begin{align}
\frac{\partial\alpha_{c}(x_{1},x_{2})}{\partial x_{1}}&=\frac{1}{c},\\
\frac{\partial\alpha_{c}(x_{1},x_{2})}{\partial x_{2}}&=0.
\end{align}
	There is no solution of $\alpha_{c}(x_{1},x_{2})$ that can be obtained from Eq. \ref{c1} and satisfies the two conditions in Eqs. \ref{p-invariance_c1} and \ref{p-invariance_c2}.
	This implies that the system is not $\mathbb{P}$-invariant in $c$.

\subsubsection{Verification of the DC property}
As a demonstration to show that $\mathbb{P}$-invariance property is more general than DC property, we used the DC property definition by Karin et al. \citep{Karin2016} in the system \ref{eq:dc_extended_control}, containing Eqs. \ref{eq:dc_extended_x1}-\ref{eq:dc_extended_y}. 
By choosing $v_1=s x_1$ and $ v_2= x_2$, for $s \neq 0$ we have:
\begin{subequations}
\begin{align}
	\label{DC_k_1}
	\dot{v}_{1} &= -cv_1\big(r(t)-v_2\big), \\
	\dot{v}_{2} &= bv_2+d(t)+v_1\big(lr(t)-v_2\big),\\
y&=v_2.
	\end{align}
\end{subequations}
Therefore, we can assume $s = 1$, which means DC property with respect to $s \neq 0$.

By choosing $v_1=x_1$ and $ v_2= b x_2$, for $s \neq 0$ we have:
\begin{subequations}
\begin{align}
	\label{DC_k_1}
	\dot{v}_{1} &= -cv_1\big(r(t)-\frac{1}{b}v_2\big), \\
	\dot{v}_{2} &= bv_2+bd(t)+bv_1\big(lr(t)-\frac{1}{b}v_2\big),\\
y&=v_2.
	\end{align}
\end{subequations}
Therefore, we cannot assume $b = 1$, which means for $b$, we could not find a transformation to prove DC property, but as DC property is a sufficient condition, we cannot claim that it is not DC property with respect to $b$. 
However, as $\mathbb{P}$-invariance property is a sufficient and necessary condition for DC, we proved that the system does not have DC for variation in $b$.

By choosing $v_1=c x_1$ and $ v_2= x_2$, for $c \neq 0$ we have:
\begin{subequations}
\begin{align}
	\label{DC_k_1}
	\dot{v}_{1} &= -cv_1\big(r(t)-v_2\big), \\
	\dot{v}_{2} &= bv_2+d(t)+\frac{s}{c}v_1\big(lr(t)-v_2\big),\\
y&=v_2.
	\end{align}
\end{subequations}
Therefore, we cannot assume $c = 1$, which means for $c$, we could not find a transformation to prove DC property, but as DC property is a sufficient condition, we cannot claim that it is not DC property with respect to $c$.
However, as $\mathbb{P}$-invariance property is a sufficient and necessary condition for DC, we proved that the system does not have DC for variation in $c$.

\section{Numerical simulation}
\label{Numerical_simulation}
In this section, we discuss and exemplify the theoretical results of our research by numerical simulations. 
We verify the phase portrait, influence of the adaptive controller, and the $\mathbb{P}$-invariance property by using step-like responses for input $r(t)$ and disturbance $d(t)$. 
To investigate the $\mathbb{P}$-invariance property with respect to the parameters $s$ and $b$, the system \ref{extend} is first brought to its equilibrium. 
Next, by perturbing the system with a step-like response $d(t)$ and changes in $s$ and $b$ separately, we check whether it returns to the equilibrium or not.

As analyzed in Section \ref{append:dc_modified_phase}, equilibrium point $E_1$ is always a saddle point, and to have stability at equilibrium point $E_2$, the main condition is $0<l<1$. Therefore, we chose initial conditions such that all solutions converge to $E_2$, i.e., a stable equilibrium point.
\begin{align}\label{parameters}
 b=0.3,\hspace{1mm} d(0)=0.01, \hspace{1mm} c=2,\hspace{1mm} r(0)=11,\hspace{1mm} l=0.7, s=0.25
\end{align}
Hence the system \ref{extend} is:
	\begin{subequations}\label{eq:extend_numeric}
	\begin{align}
		\frac{dy}{dt} &= 0.3y(t) + 0.01 + sz(t)\big(7.7-y(t)\big) , \\
		\frac{dz}{dt} &= -2z(t)\big(11-y(t)\big),
	\end{align}
	\end{subequations}
The local stability of the system can be analyzed by calculating the eigenvalues at each equilibrium point. 
 When the real parts of the eigenvalues are negative, the equilibrium point is locally stable. 

\begin{figure}[tbh]
  \begin{minipage}{\linewidth}
\begin{subfigure}[b]{0.5\linewidth}
\begin{center}
			\includegraphics[trim={0 0 0 0cm},clip,width=\columnwidth]{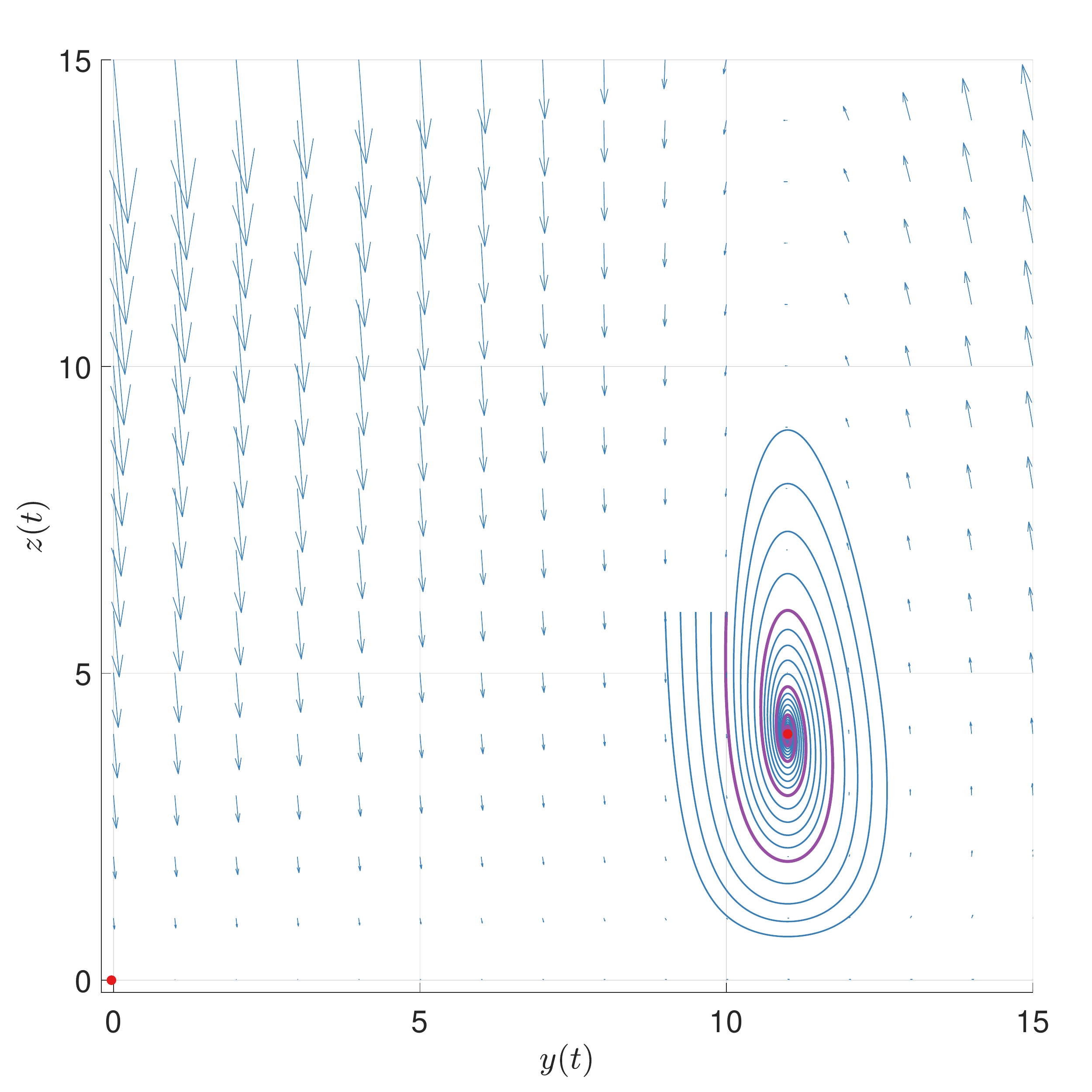}
		\subcaption{$s=0.25$, $b=0.3$, $c=2$
		\label{fig:DC_Phase_portrait_original}}
\end{center}
\end{subfigure}
     \begin{subfigure}[b]{0.5\linewidth}
\begin{center}
		\includegraphics[trim={0 0 0 0cm},clip,width=\columnwidth]{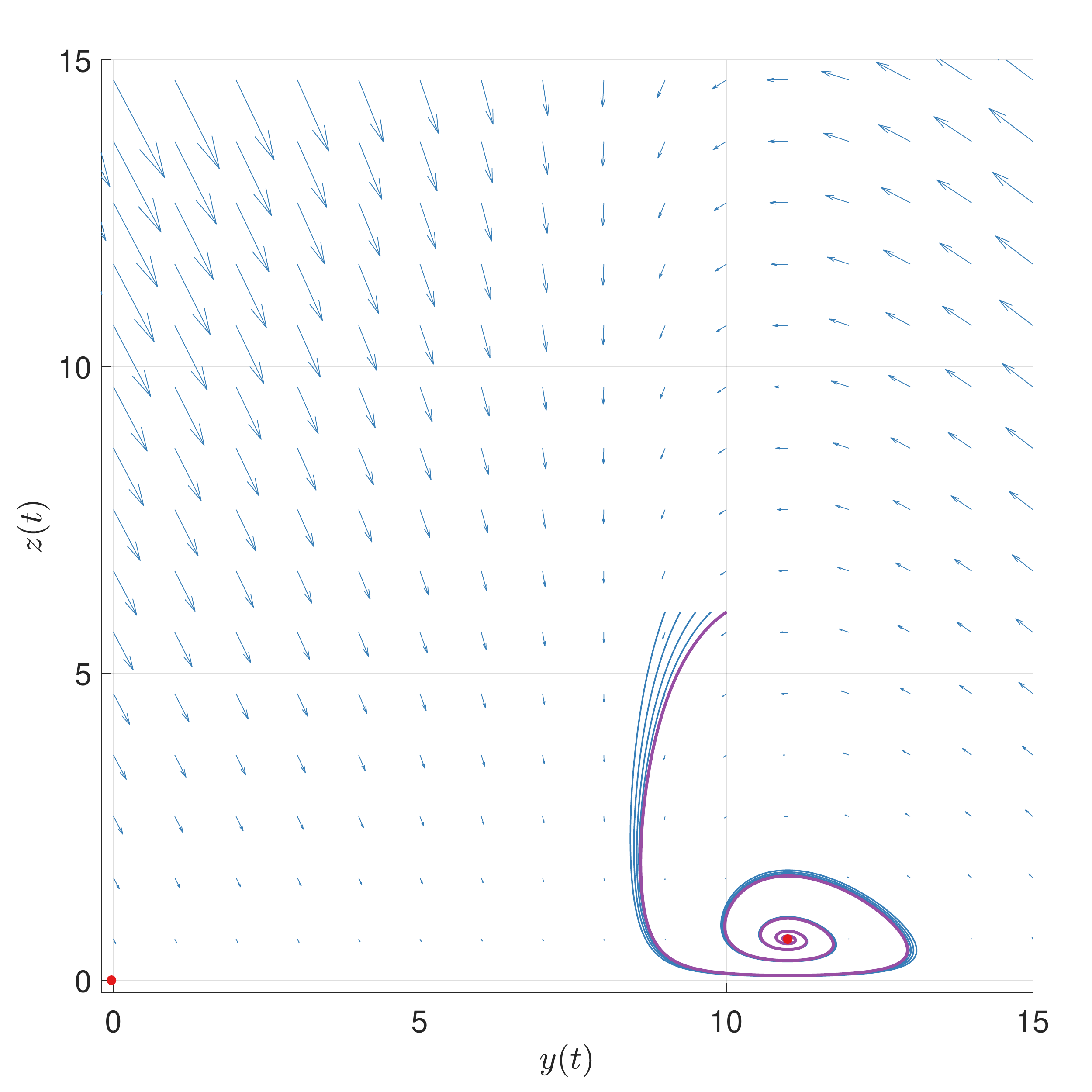}
	\subcaption{ $s=1.5$
	\label{fig:DC_Phase_portrait_s}}
\end{center}
   \end{subfigure}
     \begin{subfigure}[b]{0.5\linewidth}
\begin{center}
		\includegraphics[trim={0 0 0 0cm},clip,width=\columnwidth]{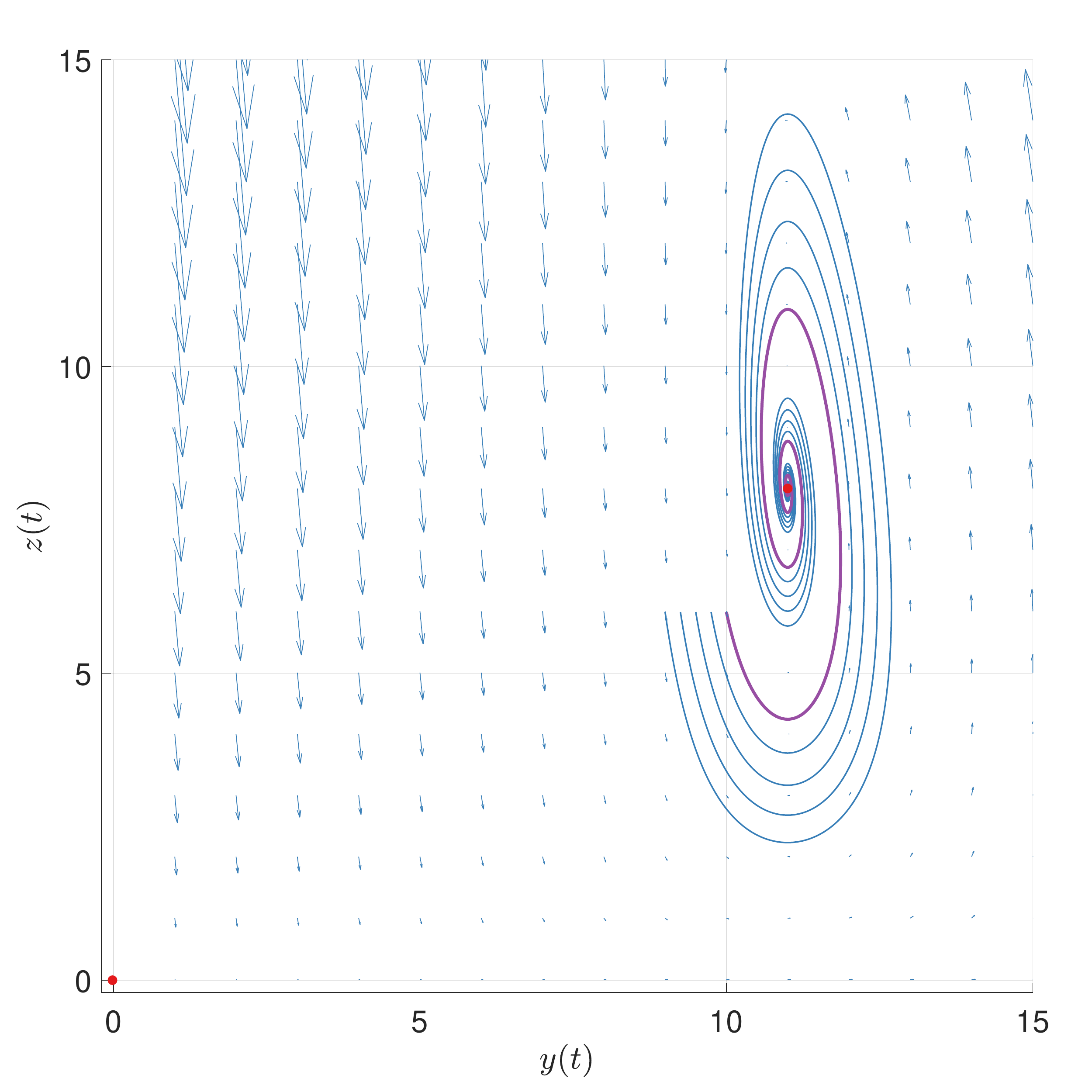}
	\subcaption{ $b=0.6$
	\label{fig:DC_Phase_portrait_b}}
\end{center}
   \end{subfigure}
     \begin{subfigure}[b]{0.5\linewidth}
\begin{center}
		\includegraphics[trim={0 0 0 0cm},clip,width=\columnwidth]{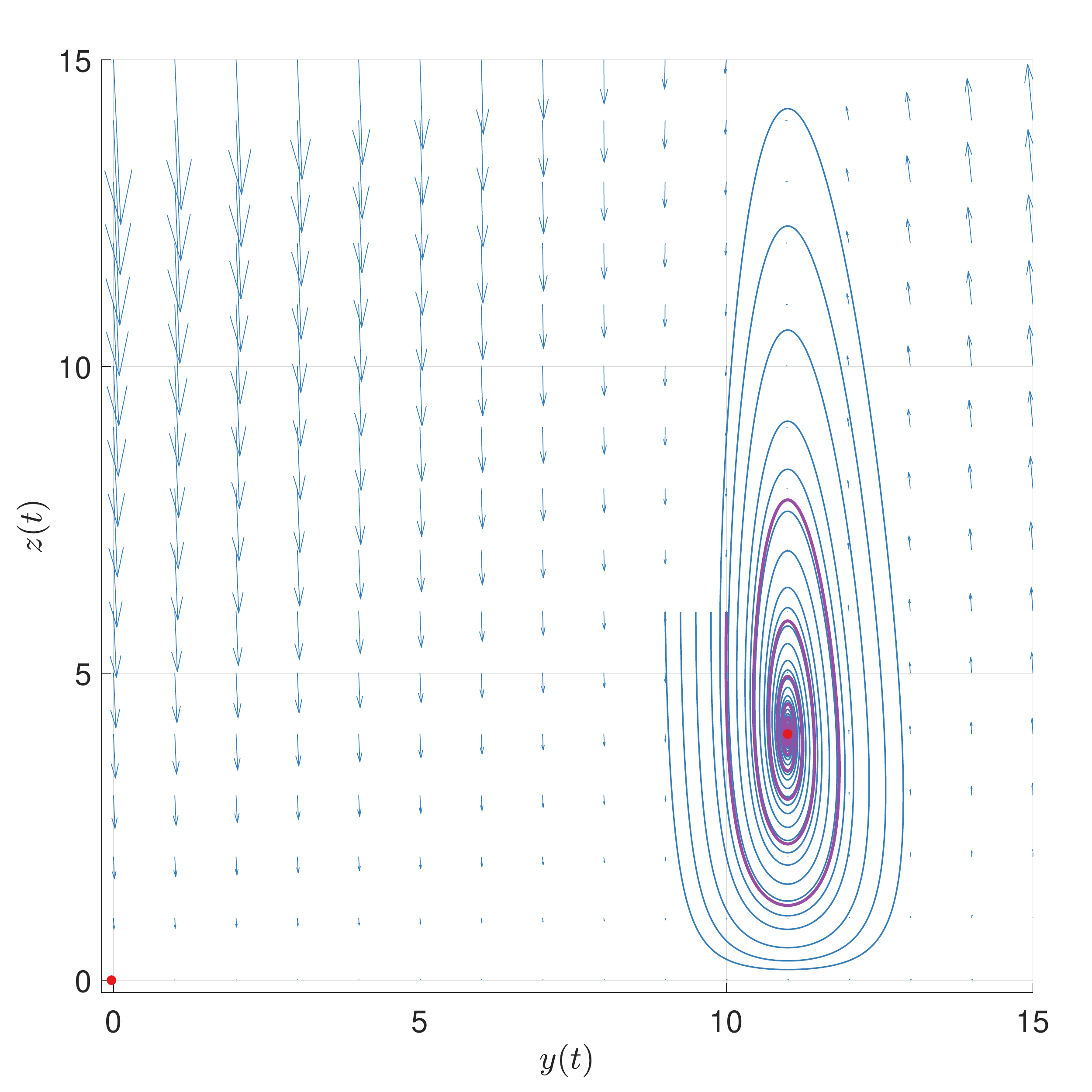}
	\subcaption{ $c=4$
	\label{fig:DC_Phase_portrait_c}}
\end{center}
   \end{subfigure}
\caption{ 
Phase portraits with different values of $s$, $b$ and $c$ show that the stable equilibrium $E_2$ has almost the same region of attraction in all cases but the trajectories differ. 
The region of attraction is determined by $E_1$.
(\subref{fig:DC_Phase_portrait_original}) The phase portrait for parameters \ref{parameters}. (\subref{fig:DC_Phase_portrait_s}) The phase portrait for parameters \ref{parameters}, except $s$ that is changed from $0.25$ to $1.5$. 
(\subref{fig:DC_Phase_portrait_b}) The phase portrait for parameters \ref{parameters}, except $b$ that is changed from $0.3$ to $0.6$.
(\subref{fig:DC_Phase_portrait_c}) The phase portrait for parameters \ref{parameters}, except $c$ that is changed from $2$ to $4$.
 \label{fig:DC_Phase_portrait}}
\end{minipage}
\end{figure}

We simulated the step-like response with initial input $r(0)=11$ and a single pulse with amplitude $5$ from time $0$ to $400$. 
We verified the results with different parameters for $s$, $b$ and $c$.
The phase portrait for the original parameters in \ref{parameters} with different values of $s$, $b$ and $c$ is shown in Fig. \ref{fig:DC_Phase_portrait}.

For $s=0.25$, two red dots in Fig. \ref{fig:DC_Phase_portrait_original} represent the equilibrium points $E_1=(-0.033, 0)$ and $E_2=(11.000, 4.012)$, with eigenvalues
\begin{align}\label{parameters_eigenvalues}
\lambda(E_1)=\{0.300, -22.067\},\hspace{5mm}
 \lambda(E_2)=\{-0.351+ 2.549 i , -0.351- 2.549 i \}.
\end{align}
Since $E_2$ is a stable equilibrium point, all trajectories in its region of attraction approach it. 
\\
If we multiply $s$ by 6 times ($s=1.5$), we obtain the equilibrium points $E_1=(-0.033, 0)$ and $E_2=(11.000, 0.669)$ in Fig. \ref{fig:DC_Phase_portrait_s}, with eigenvalues 
\begin{align}\label{parameters_eigenvalues}
\lambda(E_1)=\{0.300, -22.067\},\hspace{5mm}
 \lambda(E_2)=\{-0.351+ 2.549 i , -0.351- 2.549 i \}.
\end{align}
Again, since $E_2$ is a stable equilibrium point, all trajectories in its region of attraction approach it. 
\\
In Fig. \ref{fig:DC_Phase_portrait_b}, we choose $b=0.6$, which is twice as large as the original one, and it alters the equilibrium points to $E_1=(-0.017, 0)$ and $E_2=(11.000, 8.012)$, with eigenvalues
\begin{align}\label{parameters_eigenvalues}
\lambda(E_1)=\{0.6, -22.033\},\hspace{5mm}
 \lambda(E_2)=\{-0.701+ 3.568 i , -0.701 - 3.568 i \}.
\end{align}
Since $E_2$ is a stable equilibrium point, all trajectories in its region of attraction approach it. \\
Finally, if we multiply $c$ by two ($c=4$), we get the equilibrium points $E_1=(-0.033, 0)$ and $E_2=(11.000, 4.012)$ in Fig. \ref{fig:DC_Phase_portrait_c}, with eigenvalues 
\begin{align}\label{parameters_eigenvalues}
\lambda(E_1)=\{0.300, -44.133\},\hspace{5mm}
 \lambda(E_2)=\{-0.351+ 3.622 i , -0.351- 3.622 i \}.
\end{align}
All trajectories in the region of attraction approach $E_2$ as it is a stable equilibrium point.

\begin{figure}[tbh]
			\includegraphics[trim = 0 0 0 0, clip, width=\columnwidth]{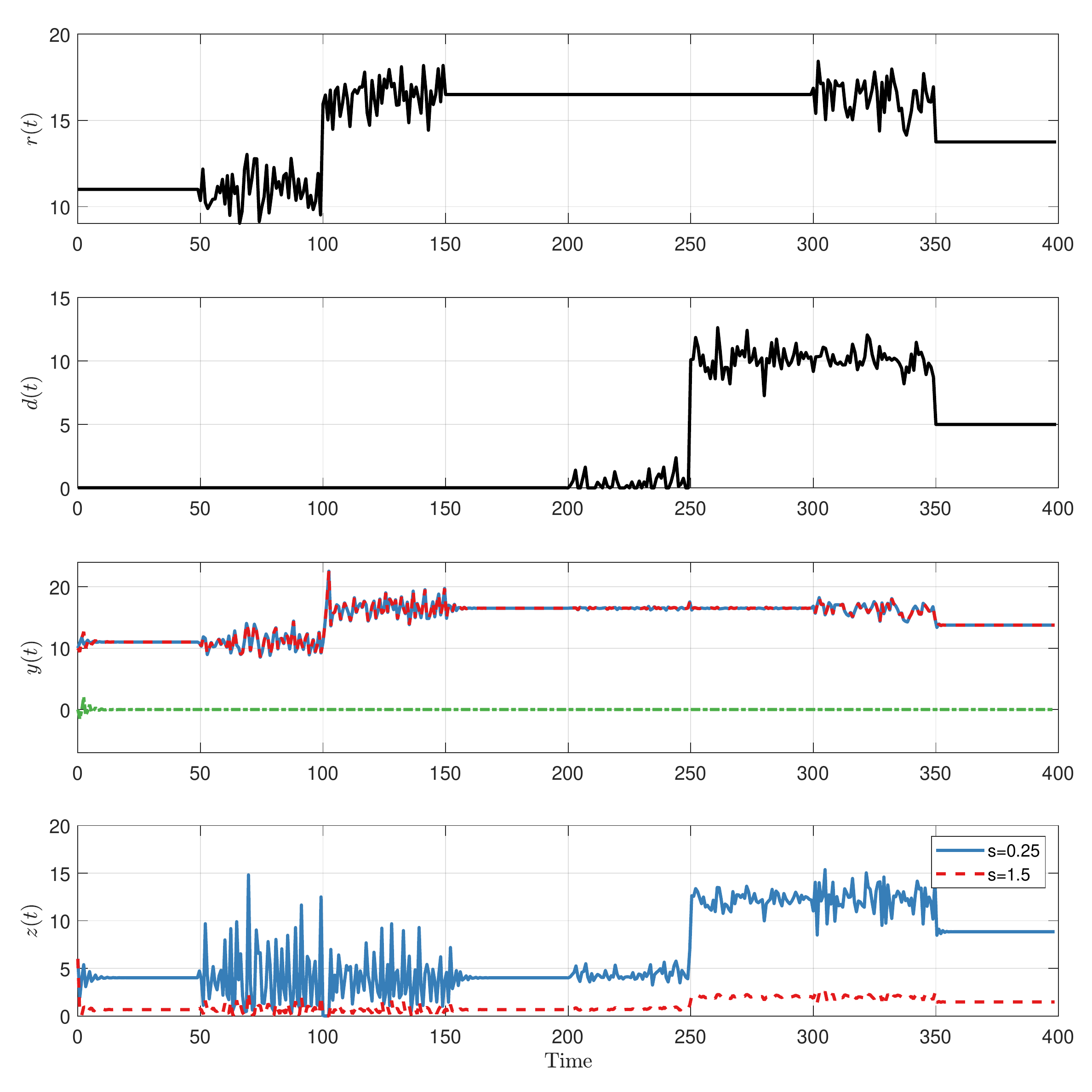}
	\caption{
Visualization of the impact of DC and lack there off on the output using time-varying step-like changes in the reference input $r(t)$ and disturbance $d(t)$ in different combinations. 
 Gaussian noise was added to the constant value of $r(t)$ and $d(t)$ during certain periods to ensure excitation. 
$y(t)$ and $z(t)$ show the comparison of the step response when $s$ is $0.25$ and $1.5$.
As we have started $y(t)$ and $z(t)$ with a distance from the equilibrium point, it takes a time to converge to the stable situation resulting at haing some residual between but then the output $y(t)$ remained identical--the residual (green dashed line) equals zero. 
A hallmark of the system being  $\mathbb{P}$-invariant with regard to $s$. 
 \label{fig:step_like_response_s}}
\end{figure}

\begin{figure}[tbh]
			\includegraphics[trim = 0 0 0 0, clip,width=\columnwidth]{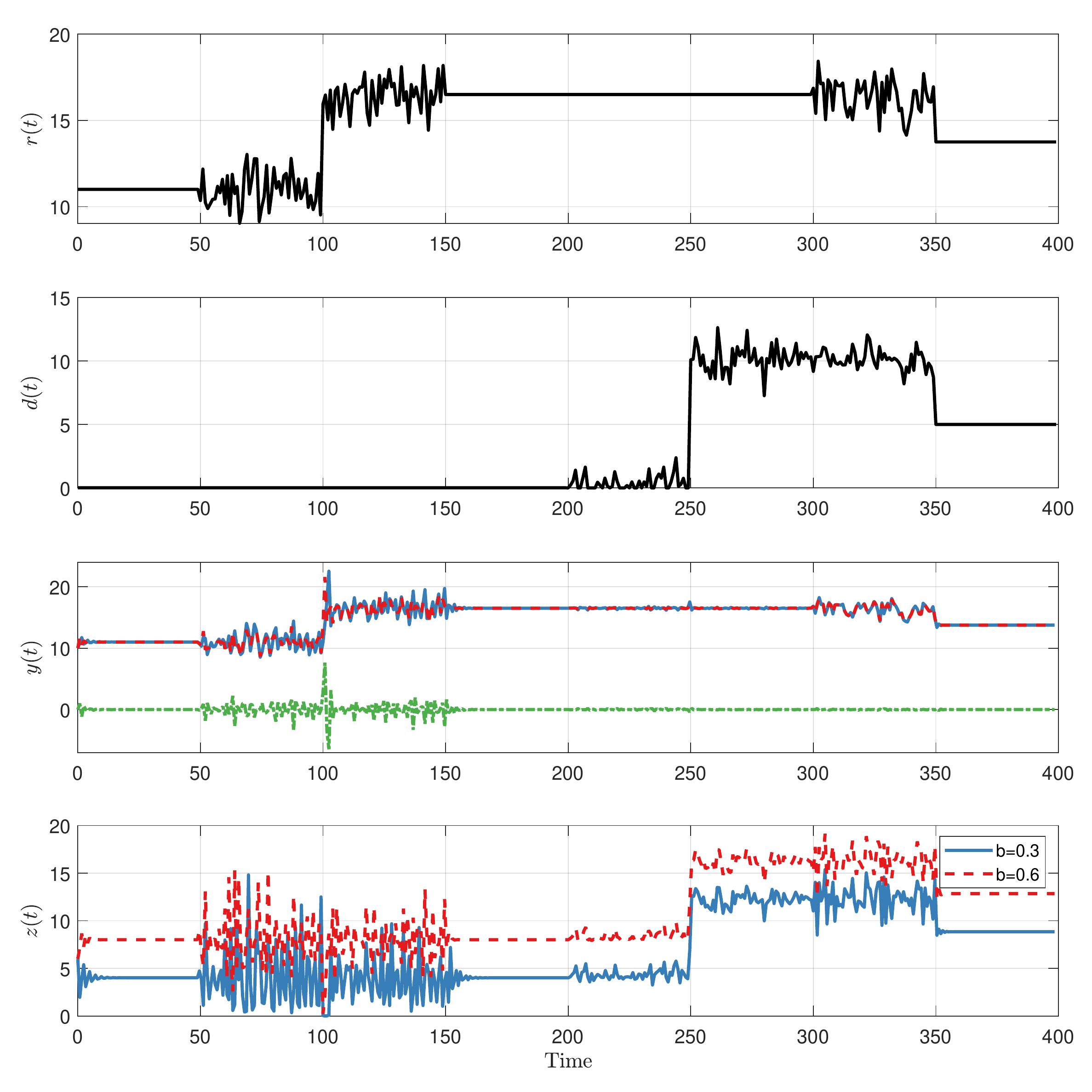}
	\caption{
Visualization of the impact of DC and lack there off on the output using time-varying step-like changes in the reference input $r(t)$ and disturbance $d(t)$ in different combinations. 
 Gaussian noise was added to the constant value of $r(t)$ and $d(t)$ during certain periods to ensure excitation. 
$y(t)$ and $z(t)$ show the comparison of the step response when $b$ is $0.3$ and $0.6$.
The output $y(t)$ differs and the residual (green dashed line) is non-zero. 
A hallmark of the system not being $\mathbb{P}$-invariant with regard to $b$. \label{fig:step_like_response_b}}
\end{figure}

\begin{figure}[tbh]
			\includegraphics[trim = 0 0 0 0, clip,width=\columnwidth]{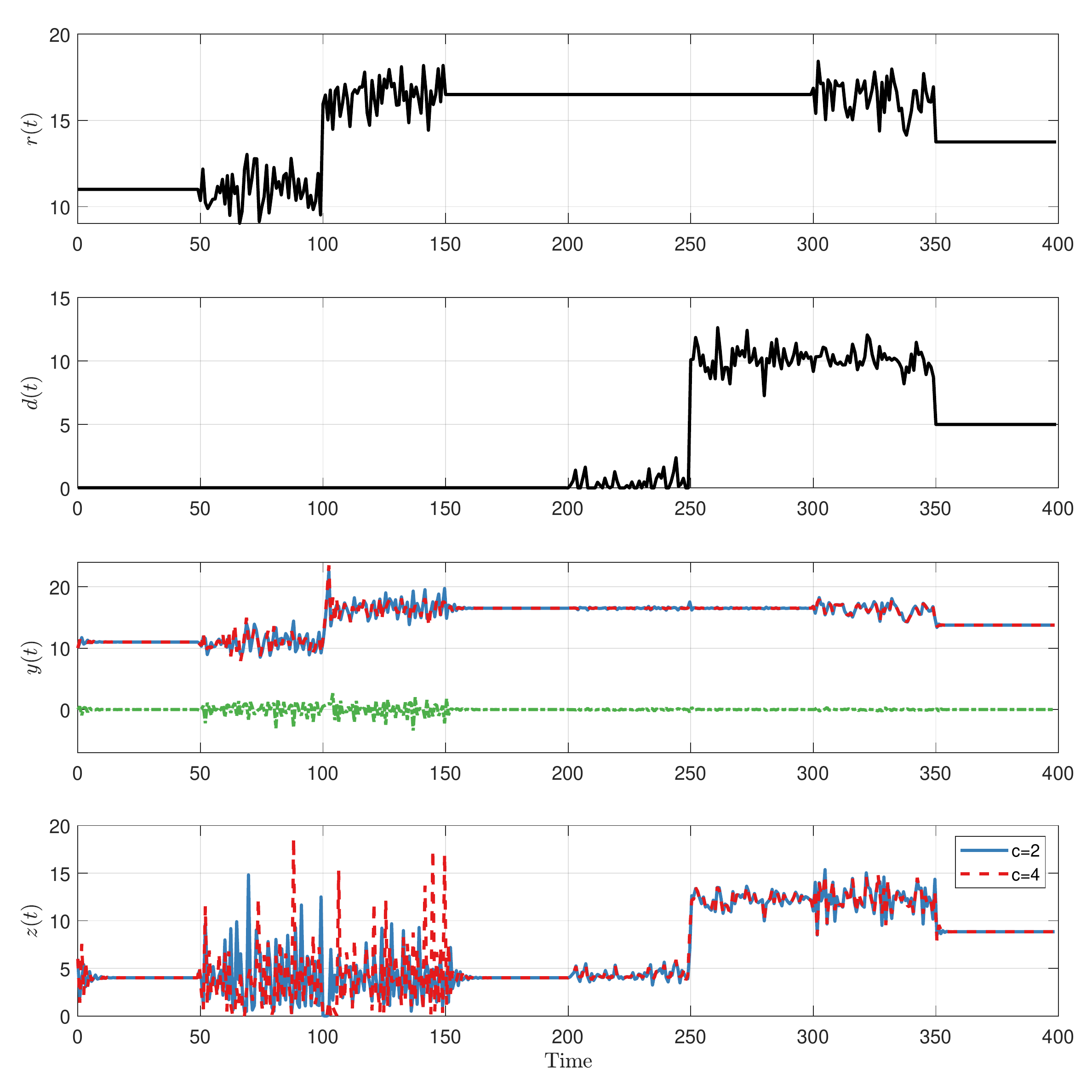}
	\caption{
Visualization of the impact of DC and lack there off on the output using time-varying step-like changes in the reference input $r(t)$ and disturbance $d(t)$ in different combinations. 
 Gaussian noise was added to the constant value of $r(t)$ and $d(t)$ during certain periods to ensure excitation. 
$y(t)$ and $z(t)$ show the comparison of the step response when $c$ is $2$ and $4$.
The output $y(t)$ differs and the residual (green dashed line) is non-zero. 
A hallmark of the system not being $\mathbb{P}$-invariant with regard to $c$.\label{fig:step_like_response_c}}
\end{figure}

After verifying the stability of the system, we investigated the $\mathbb{P}$-invariance property under different situations.
In all Figs. \ref{fig:step_like_response_s}, \ref{fig:step_like_response_b} and \ref{fig:step_like_response_c} $r(t)$ and $d(t)$ are the same and show the time-varying step-like response of the reference input $r(t)$ and disturbance $d(t)$
These inputs were also subject to additional noise from a standard normal distribution.
We tested different combinations of the reference $r(t)$ and disturbance $d(t)$ to exemplify the $\mathbb{P}$-invariance property under different scenarios. 
Both the input $r(t)$ and the disturbance $d(t)$ began with the starting values defined at \ref{parameters} and remained constant from time 0 until time 50.
The input $r(t)$ changes while disturbance $d(t)$ remains constant when time is between 50 and 150. 
Both the input $r(t)$ and disturbance $d(t)$ remain constant between time 150 and 200. 
In the time interval ($200, 300$), the disturbance $d(t)$ changes while the input $r(t)$ remains constant. 
When time is between ($300, 350$), both the input $r(t)$ and disturbance $d(t)$ change. 
Finally, both converge to a new amount ($r=13.75, d=5$) and remain constant in the time interval ($350, 400$). 

The purple trajectories in Fig. \ref{fig:DC_Phase_portrait} illustrates the starting position in the Figs. \ref{fig:step_like_response_s}, \ref{fig:step_like_response_b} and \ref{fig:step_like_response_c}. 
As a result, we are in a stable situation at the start, however it may take some time to achieve stability.
In all Figs. \ref{fig:step_like_response_s}, \ref{fig:step_like_response_b} and \ref{fig:step_like_response_c} $z(t)$ and $y(t)$ show the responses to the input and disturbance by $r(t)$ and $d(t)$. 
The green dashed line represents the residual of changes in $y(t)$, which remains zero when $s$ changes, but is non-zero when $b$ or $c$ changes. 
This is a consequence of the system having $\mathbb{P}$-invariance for parameter $s$, but not for $b$ and $c$.

\section{Conclusions}\label{sec:conclusion}
Our two state simplified and extended model based on Karin et al.'s work \cite{Karin2016} preserves the DC property when the parameter $s$ is changed. 
We have demonstrated this using the $\mathbb{P}$-invariance definition by Sontag \cite{Sontag2017}. 
With this approach, we have also shown no DC for the parameters $b$ and $c$, because the definition of $\mathbb{P}$-invariance is both sufficient and necessary.

Our example system is an exponential growth system with an adaptive proportional integral controller.
Exponential growth is a common feature of many physical systems, such as early stage of cell growth or disease spread. 
We have shown that our adaptive proportional integral feedback with DC in the control parameters $s$ can stabilize the system and ensure that the response tracks the reference input despite variation in the control parameters. 
The downside of this is that the closed loop systems behavior cannot be tuned by changing the gain of the controller as customary in e.g. PID-controllers. 
Moreover, we have demonstrated the stability of the system under a variety of conditions and plotted the phase portrait for a representative example. 
In summary, we have demonstrated an adaptive controller with $\mathbb{P}$-invariance in it's parameter $s$.
This can be beneficial for designing robust controllers that can handle environmental fluctuations, in particular in Synthetic biology, as well as for understanding biological systems during modeling and analyzing.


\section{Data availability}
All data used is included in this article.
\section{Acknowledgement}
The authors gratefully acknowledge valuable comments by Prof. Filippo Menolascina from the University of Edinburgh, UK. 
\section{Funding}
We would like to thank the Ministry of Science and Technology in Taiwan for their financial support (grants number MOST 105-2218-E-006-016-MY2, 105-2911-I-006-518, 107-2634-F-006-009, 110-2222-E-006-010, and 111-2221-E-006-186).
\section{Conflict of interest}
The authors have no conflicts of interest to declare.
\section{Declaration of Generative AI and AI-assisted technologies in the writing process}
During the preparation of this work the authors used GPT3.5 and GPT4 by OpenAI, L.L.C. in order to help us improve the readability and language of this article. After using this tool/service, the authors reviewed and edited the content as needed and take full responsibility for the content of the publication.




\clearpage
\bibliographystyle{elsarticle-num} 

\bibliography{../../LibraryAllReferences}


%
%
%
\end{document}